# Temperature dependent elastic constants for crystals with arbitrary symmetry: combined first principles and continuum elasticity theory

**Tianjiao Shao[a], Bin Wen[b, 1], Roderick Melnik[c], Shan Yao[a], Yoshiyuki Kawazoe[d], Yongjun Tian[b]**

[a]School of Materials Science and Engineering, Dalian University of Technology, Dalian 116023, China

[b]State Key Laboratory of Metastable Materials Science and Technology, Yanshan University, Qinhuangdao 066004, China

[c]M$^2$NeT Lab, Wilfrid Laurier University, Waterloo,75 University Ave. West, Ontario, Canada N2L 3C5

[d]Institute for Materials Research, Tohoku University, 2-1-1 Katahira, Aoba-ku, Sendai 980-8577, Japan

**Abstract** To study temperature dependent elastic constants, a new computational method is proposed by combining continuum elasticity theory and first principles calculations. A Gibbs free energy function with one variable with respect to strain at given temperature and pressure was derived, hence the full minimization of the Gibbs free energy with respect to temperature and lattice parameters can be put into effective operation by using first principles. Therefore, with this new theory, anisotropic thermal expansion and temperature dependent elastic constants can be obtained for crystals with arbitrary symmetry. In addition, we apply our method to hexagonal beryllium, hexagonal diamond and cubic diamond to illustrate its general applicability.

[1] Authors to whom any correspondence should be addressed
E-mail address: wenbin@ysu.edu.cn (Bin Wen), Tel: 086-335-8568761

## I. Introduction

Due to a wide range of industrial applications of high temperature alloys [1], studies of earth structure and earthquake mechanisms, as well as other applications [2-3], temperature dependent elastic constants (TDEC) of materials have long been an important subject for both experimental and theoretical investigations [4, 5]. Several experimental methods have already been developed to measure TDEC [6]. However, it is difficult to carry out experimental TDEC measurements due to complexity of sample preparation and maintaining the sample under high temperature conditions. Therefore, a number of theoretical methods have also been proposed to study the TDEC of materials [7-22].

In the 1970s, based on the analysis of temperature dependent 57 elastic constants for 22 substances, an empirical TDEC formula, now known as the Varshni equation, was proposed by Varshni [7]. Although this empirical formula can describe TDEC for crystals with arbitrary symmetry, its applicability was limited to some materials only due to the fact that two fitting parameters from experimental data were needed. After that, many similar empirical TDEC formulas have been suggested [8-10]. In 1975, TDEC for cubic crystals have been derived on the basis of a quasiharmonic anisotropic continuum model by Garber et al [11]. In 1997, a method to fully minimize the Gibbs energy with respect to lattice parameters for arbitrary symmetry crystals was developed by Taylor et al. within a quasi harmonic approximation (QHA) [12], and then TDEC of crystals with arbitrary symmetry cannot be calculated based on Taylor's model. In 2004, on the basis of the QHA and Debye model, a non-equilibrium Gibbs energy as a function of volume at given temperature was derived, providing a way to obtain an equilibrium volume at a given temperature [13-16]. Since volume is a function of only lattice parameter $a$ for cubic cystals, the lattice parameters at a given temperature can be obtained accurately, and hence an accurate TDEC for cubic crystals can be obtained by using Blanco's method [17-22]. However, since for low symmetry crystals, volume is a multiple variable function, the geometry optimization for low symmetry crystals at given temperature is currently infeasible due to the difficulty in global minimization of multiple variable functions. Therefore, TDEC can not be accurately obtained by using Blanco's method. To date, a simple and accurate method for predicting TDEC of crystals with arbitrary symmetry is still lacking. To fill this gap, in this work we propose a new method on the basis of continuum elasticity theory and first principles. A Gibbs energy function with one variable with respect to strain at given temperature and pressure was derived based on Wang's result [23], hence the geometry optimization for arbitrary symmetry crystals at given temperature can be put into effective operation. Therefore, based on the new theory developed here, both anisotropic thermal expansion coefficients and TDEC can be obtained for crystals with arbitrary symmetry. In addition, we apply our method to hexagonal beryllium, hexagonal diamond and cubic diamond to illustrate its general applicability.

## II. Theory and methodology

### A. Geometry optimization of crystals with arbitrary symmetry at given temperature and pressure

According to crystallographic theory [24], for an arbitrary crystal configuration tensor $x$, it includes the lattice lengths, lattice angles of this crystal, all free crystallographic coordinates of the atoms in non-fixed Wyckoff positions and so forth. If the crystallographic coordinates of the atoms are fixed, the crystal configuration tensor $x$ is a function of lattice lengths $a$, $b$, $c$ and lattice angles $\alpha$, $\beta$ and $\gamma$. Based on standard thermodynamics arguments [25, 26], if the system is held at fixed temperature $T$ and constant hydrostatic pressure $p$, the non-equilibrium Gibbs energy for a specific phase can be written as

$$\begin{aligned} G[x(a,b,c,\alpha,\beta,\gamma);p,T] = E[x(a,b,c,\alpha,\beta,\gamma)] + pV[x(a,b,c,\alpha,\beta,\gamma)] \\ + A_{vib}[x(a,b,c,\alpha,\beta,\gamma);T]. \end{aligned} \tag{1}$$

To obtain the Gibbs energy at equilibrium state, a full minimization of non-equilibrium Gibbs energy function (1) is needed. Due to the difficulty in global minimization of multiple variable functions, the full minimization of non-equilibrium Gibbs energy function (1) is currently infeasible by using first principles method. To address this challenge, a group of Gibbs energy functions with one deformation strain $\xi$ was derived in this paper.

Based on linear algebra and tensor analysis ideas [27], a deformed configuration tensor $X(a,b,c,\alpha,\beta,\gamma)$ can be expressed as a product of 9-dimentional deformation tensor and an initial configuration tensor, and therefore the deformed configuration tensor can be expressed as

$$X = X_0 \begin{pmatrix} e_1 & e_2 & e_3 \\ e_4 & e_5 & e_6 \\ e_7 & e_8 & e_9 \end{pmatrix} \begin{pmatrix} \xi+1 & 0 & 0 \\ 0 & \xi+1 & 0 \\ 0 & 0 & \xi+1 \end{pmatrix}, \tag{2}$$

where $X_0$ is the initial configuration tensor, $\begin{pmatrix} e_1 & e_2 & e_3 \\ e_4 & e_5 & e_6 \\ e_7 & e_8 & e_9 \end{pmatrix}$ is a normal deformation mode tensor, and $\xi$ is deformation strain.

If the deformation is symmetric, the deformed configuration tensor can be rewritten as

$$X = X_0 \begin{pmatrix} e_1 & e_2 & e_3 \\ e_2 & e_4 & e_5 \\ e_3 & e_5 & e_6 \end{pmatrix} \begin{pmatrix} \xi+1 & 0 & 0 \\ 0 & \xi+1 & 0 \\ 0 & 0 & \xi+1 \end{pmatrix}. \tag{3}$$

If the deformation mode is fixed, the deformed configuration tensor is a function of only deformation strain $\xi$. Therefore, the non-equilibtium Gibbs energy can be expressed as

$$G[X(a,b,c,\alpha,\beta,\gamma); p,T] = E[X(\xi)] + pV[X(\xi)] + A_{vib}[X(\xi);T], \tag{4}$$

where $E[X(\xi)]$ is the total energy of the specific deforming configuration. $A_{vib}[X(\xi);T]$ is the vibrational Helmholtz free energy, which can be calculated from phonon density of states by using the QHA,

$$A_{vib}[X(\xi);T] = \int_0^\infty [\frac{1}{2}\hbar\omega + kT\ln(1-\mathrm{e}^{-\hbar\omega/k_BT})]g[X(\xi);\omega]d\omega, \tag{5}$$

where $\omega$ is the phonon frequency, and $T$ is the temperature, $k$ and $\hbar$ are the Boltzmann constant and the reduced Planck constant, respectively. At the same time, the vibrational Helmholtz free energy $A_{vib}[X(\xi);T]$ also can be calculated by using first principles molecular dynamic (FPMD) calculations.

Based on Wang's work [23], if the deformation tensor is symmetric, the Gibbs energy for an equilibrium state at given temperature $T$ and pressure $p$ can be obtained by minimizing the formula (4) with respect to variable $\xi$,

$$G^*(\xi^0_{p,T}; p,T) = \min_{\xi}\left\{E[X(\xi)] + pV[X(\xi)] + A_{vib}[X(\xi);T]\right\}, \tag{6}$$

and the corresponding equilibrium strain $\xi^0_{p,T}$ at the given temperature $T$ and pressure $p$ can be obtained. Then, for a fixed deformation mode, equilibrium strain $\xi^0_{p,T}$ is the function of configuration factor $(a,b,c,\alpha,\beta,\gamma)$. Therefore, to obtain the equilibrium configuration at a given temperature $T$ and pressure $p$, a group of six independent deformation modes are needed, and it is expressed formally as follows:

$$\begin{cases} F_1(a,b,c,\alpha,\beta,\gamma,\xi^0_{1,p,T}) = 0 \\ F_2(a,b,c,\alpha,\beta,\gamma,\xi^0_{2,p,T}) = 0 \\ F_3(a,b,c,\alpha,\beta,\gamma,\xi^0_{3,p,T}) = 0 \\ F_4(a,b,c,\alpha,\beta,\gamma,\xi^0_{4,p,T}) = 0 \\ F_5(a,b,c,\alpha,\beta,\gamma,\xi^0_{5,p,T}) = 0 \\ F_6(a,b,c,\alpha,\beta,\gamma,\xi^0_{6,p,T}) = 0 \end{cases}. \quad (7)$$

By solving system of equations (7), the lattice parameters can be determined, and in addition, the thermal expansion coefficients can be obtained.

For a high symmetry lattice, the dimension of configuration tensor $X(a,b,c,\alpha,\beta,\gamma)$ can be reduced to a dimension less than six, and therefore the number of functions in system (7) can be reduced. For example, for the cubic crystal, the dimension number of configuration tensor $X(a,b,c,\alpha,\beta,\gamma)$ is one, therefore, we can use just one function to obtain the lattice parameter of cubic crystal. In this work, we choose $(\xi_1,\xi_1,\xi_1,0,0,0)$ as the deformation tensor, and in this case the Eq. (6) can be written as following,

$$G^*(\xi^0_{1,p,T};p,T) = \min_{\xi_1}[E(\xi_1) + pV(\xi_1) + A_{vib}(\xi_1;T)]. \quad (8)$$

By soloving Eq. (8), the equilibrium deformation strain $\xi^0_{1,p,T}$ at given temperature $T$ and pressure $p$ can be obtained, and the Eq. (7) can be simplified as

$$a_{p,T} = \xi^0_{1,p,T} a_0, \quad (9)$$

where $a_0$ represents the optimized lattice parameter at 0 K obtained by first principles geometry optimization processing directly; $a_{p,T}$ represents the equilibrium lattice at given temperature $T$ and pressure $p$, and it is the final optimized lattice parameter for cubic crystal. Hence, by solving Eq. (9), the optimized lattice parameter $a_{p,T}$ at given temperature $T$ and pressure $p$ can be obtained. Based on these temperature dependent lattice parameters, the thermal expansions coefficient can also be obtained.

For hexagonal crystals, we have two independent geometry configuration variables, so that the dimension number of configuration tensor $X(a,b,c,\alpha,\beta,\gamma)$ is 2. Therefore, two functions are needed to obtain the hexagonal crystal lattice parameters. In this work, we choose $(\xi_1,0,0,0,0,0)$ and $(0,0,\xi_2,0,0,0)$ as the deformation tensors, therefore, the Eq. (6) can be written as the following system of equations:

$$\begin{cases} G^*(\xi^0_{1,p,T};p,T)=\min\limits_{\xi_1}[E(\xi_1)+pV(\xi_1)+A_{vib}(\xi_1;T)] \\ G^*(\xi^0_{2,p,T};p,T)=\min\limits_{\xi_2}[E(\xi_2)+pV(\xi_2)+A_{vib}(\xi_2;T)] \end{cases}. \tag{10}$$

Further, the Eq. (7) can be simplified as:

$$\begin{cases} a_{p,T} = \xi^0_{1,p,T}a_0 \\ c_{p,T} = \xi^0_{2,p,T}c_0 \end{cases}, \tag{11}$$

where $a_0$ and $c_0$ represent the optimized lattice parameters for hexagonal crystal at 0 K by first principles geometry optimization processing directly; $a_{p,T}$ and $c_{p,T}$ represent the equilibrium lattice parameters for $a$ and $c$ directions at given temperature $T$ and pressure $p$, respectively. Finally, temperature dependent lattice parameters and the corresponding thermal expansions coefficient for $a$ and $c$ directions can be obtained.

**B. Temperature dependent elastic constants for crystals with arbitrary symmetry**

After optimized crystal lattice parameters at given temperature and pressure were obtained, the TDEC can be studied. On the basis of continuum elasticity theory [28, 29], elastic constants can be considered as strain derivatives of the Helmholtz free energy. Following the ideas of Section A, if the deformation tensor is symmetric and fixed, the Helmholtz free energy $F[X(\xi);p,T]$ can be expressed as

$$F[X(\xi);p,T]=E[X(\xi)]+A_{vib}[X(\xi);T], \tag{12}$$

where $E[X(\xi)]$ is the total energy of the specific deforming configuration. $A_{vib}[X(\xi);T]$ is the vibrational Helmholtz free energy, which can be calculated from phonon density of states by using the QHA or by FPMD simulations.

Then, for different deformation modes, strain derivatives of the Helmholtz free energy correspond to the different elastic constants or their linear combinations. Therefore, in order to investigate elastic constants in detailds, a series of the Helmholtz free energy $F[X(\xi); p,T]$ curves with respect to strain $\xi$ at given temperature $T$ and pressure $p$ have to be evaluated. For *P1* crystal, a function group of 21 independent deformation modes and the corresponding strain derivatives of the Helmholtz free energy are needed, and it is expressed formally as follows:

$$\begin{cases} F_1(D_1^{p,T}, C_{11}^{p,T}, C_{12}^{p,T}, C_{22}^{p,T} \cdots, C_{ij}^{p,T}, \cdots, C_{66}^{p,T}) = 0 \\ F_2(D_2^{p,T}, C_{11}^{p,T}, C_{12}^{p,T}, C_{22}^{p,T} \cdots, C_{ij}^{p,T}, \cdots, C_{66}^{p,T}) = 0 \\ \vdots \\ F_i(D_i^{p,T}, C_{11}^{p,T}, C_{12}^{p,T}, C_{22}^{p,T} \cdots, C_{ij}^{p,T}, \cdots, C_{66}^{p,T}) = 0 \\ \vdots \\ F_{21}(D_{21}^{p,T}, \underbrace{C_{11}^{p,T}, C_{12}^{p,T}, C_{22}^{p,T} \cdots, C_{ij}^{p,T}, \cdots, C_{66}^{p,T}}_{21}) = 0 \end{cases}, \tag{13}$$

where $D_1^{p,T} \cdots D_i^{p,T} \cdots D_{21}^{p,T}$ represent the second-order strain derivatives of the Helmholtz free energy under different deformation modes, $C_{11}^{p,T} \cdots C_{ij}^{p,T} \cdots C_{66}^{p,T}$ represent the elastic constants at given temperature and pressure.

For cubic crystals, there three independent elastic constants, therefore, the Eq. (13) can be reduced to a function group with three independent functions. For example, if we choose $(\xi, -\xi, \xi^2/(1-\xi^2), 0, 0, 0)$ , $(\xi^2/(1-\xi^2), 0, 0, 2\xi, 0, 0)$ and $(\xi, \xi, \xi, 0, 0, 0)$ as three deformation modes, the function group Eq. (13) can be written as follows,

$$\begin{cases} C_{11}^{p,T} - C_{12}^{p,T} = D_1^{p,T} \\ 4C_{44}^{p,T} = D_2^{p,T} \\ 3C_{11}^{p,T} + 6C_{12}^{p,T} = D_3^{p,T} \end{cases}, \tag{14}$$

where $D_1^{p,T}$, $D_2^{p,T}$ and $D_3^{p,T}$ represent second-order strain derivatives of the Helmholtz free energy under the above three deformation modes, respectively. $C_{11}^{p,T}$, $C_{12}^{p,T}$ and $C_{44}^{p,T}$ represent the elastic constants for cubic crystals at given temperature and pressure.

By solving the linearly function group (14), the three independent elastic constants for the cubic lattice at given temperature and pressure can be obtained.

Since for hexagonal crystals, there are five independent elastic constants, therefore, the Eq. (13) can be reduced to a function group with five independent functions. For example, if we choose $(\xi-1,-\xi+1,(\xi-1)^2/(1-(\xi-1)^2),0,0,0)$, $((\xi-1),(\xi-1)^2/(1-(\xi-1)^2),-(\xi-1),0,0,0)$, $((\xi-1)^2/(1-(\xi-1)^2),0,0,2(\xi-1),0,0)$, $((\xi-1),(\xi-1),0,0,0,0)$ and $((\xi-1),(\xi-1),(\xi-1),0,0,0)$ as five deformation modes, the function group Eq. (13) can be written as follows

$$\begin{cases} 2C_{11}{}^{p,T} - 2C_{12}{}^{p,T} = D_1{}^{p,T} \\ C_{11}{}^{p,T} + C_{33}{}^{p,T} - 2C_{13}{}^{p,T} = D_2{}^{p,T} \\ 4C_{44}{}^{p,T} = D_3{}^{p,T} \\ 2C_{11}{}^{p,T} + 2C_{12}{}^{p,T} = D_4{}^{p,T} \\ 2C_{11}{}^{p,T} + 4C_{12}{}^{p,T} + 2C_{13}{}^{p,T} + C_{33}{}^{p,T} = D_5{}^{p,T} \end{cases}, \quad (15)$$

where $D_1^{p,T}$, $D_2^{p,T}$, $D_3^{p,T}$, $D_4^{p,T}$ and $D_5^{p,T}$ represent second-order strain derivatives of the Helmholtz free energy under the above five kinds of deformation modes, respectively. $C_{11}^{p,T}$, $C_{12}^{p,T}$, $C_{13}^{p,T}$, $C_{33}^{p,T}$ and $C_{44}^{p,T}$ represent the elastic constants for hexagonal crystals at given temperature and pressure.

Similarly, by solving the system of functions (15), the five independent elastic constants for the hexagonal lattice at given temperature and pressure can be obtained.

**C. Computational method**

To validate our method and to provide further details of our computational steps,

hexagonal beryllium, hexagonal diamond and cubic diamond have been tested. In this work, all the first principles calculations have been performed by using the Vienna ab initio simulation package (VASP) developed by the Hafner Research Group [30]. Computations have been performed by using the density functional theory within the local density approximation (LDA) [31, 32], while the ion-electron interaction has been modeled by the projector-augmented wave (PAW) method [33]. A plane-wave energy cutoff of 600 eV has been used and the Brillouin zone of the cells has been sampled by $14 \times 14 \times 8$ **k**-point mesh for hexagonal beryllium, $10 \times 10 \times 8$ **k**-point mesh for hexagonal diamond, $8 \times 8 \times 8$ **k**-point mesh for cubic diamond. The $2 \times 2 \times 2$ supercells have been used in this work, and the atom numbers in these supercells have been 16, 32 and 64 for hexagonal beryllium, hexagonal diamond and cubic diamond, respectively.

To calculate the vibrational Helmholtz free energy, both QHA and FPMD methods have been used in this work. The QHA calculation has been performed by combing VASP and PHONOPY codes [34, 35], and the FPMD method has been performed by using the VASP package. For the QHA calculations performed by the PHONOPY code [34, 35], a supercell method with finite displacement [36] has been employed, and a supercell containing $2 \times 2 \times 2$ unit cells has used in this work. For the FPMD performed for cubic diamond, a $2 \times 2 \times 2$ cubic diamond unit cell with 64 atoms has been used; cutoff energy and k-point sampling have been used as 600 eV and Gama point, respectively. The canonical ensemble (NVT) which conserves moles (N), volume (V) and temperature (T) with the Nose thermostat has been used. The simulation time is 5 ps and the time step is 0.5 fs.

## III. Applications

### A. Hexagonal beryllium

Since beryllium has hexagonal symmetry, the temperature dependent lattice parameters and thermal expansion coefficients can be obtained by using Eq. (10) and (11), and the TDEC can be studied by using Eq. (15). In detail, five deformation tensors were used, and they are listed in Table 1. By using the QHA method, five Gibbs energy densities against these strain $\xi$ curves were calculated at different temperature conditions and they are shown in Figure 1a-f. For fixed deformation mode, the Gibbs energy is a function of one variable $\xi$ at a given temperature; hence the equilibrium Gibbs energy at a given temperature can be obtained easily by minimizing the Gibbs energy function with respect to variable $\xi$. For a fixed crystal, hexagonal beryllium, both the equilibrium crystal configuration and the corresponding Gibbs energy at a given temperature are unitary. Figure 1k shows the equilibrium lattice Gibbs energy against temperature with respect to five deformation modes. As can be seen from Figure 1k, these curves coincide with each other. This result implies high accuracy of our present computational method. At the same time, temperature dependent lattice parameters and thermal expansion coefficients can be obtained and they are shown in Figure 1m and Figure 2a, respectively. In particular, as can be seen from Figure 2a, the calculated thermal expansion coefficients agree well with the corresponding experimental ones [37]. Due to the thermal expansion effect, the strain and volume for equilibrium lattice are varying with temperature. Because the pressure is zero in this study, the Gibbs energy equals to the Helmholtz free energy. Therefore, relationships for the Helmholtz free energy densities with respect to normal strain $\zeta$ can be obtained by recalculating the relationships between Gibbs energy densities with respect to strain, and they are shown in Figures 1f-j. By polynomial fitting these Helmholtz free energy densities to strain (see in Figures 1f-j), the second order coefficient can be obtained, and it equals to the corresponding linear combination of elastic constants (as listed in Table 1). The elastic constants at a given temperature can be calculated by solving simultaneously these five linear equations according to the Eq. (15). Figure 2b shows the comparison of calculated and experimental TDEC. As can be seen, our calculated results agree well with experimental ones [38, 39]. This further confirms high accuracy of the developed computational method. As can be seen from Figure 2b, the calculated elastic constants are slightly larger than the corresponding experimental values and the slopes of experimental elastic constants with respect to temperature are slightly larger than the calculated slopes. For real materials under high temperature, due to configuration entropy, the existence of defects is inevitable, and the defects concentration is increasing with increasing temperature. Therefore, the experimental elastic constants are slightly smaller than calculated values and the slopes of experimental elastic constants with respect to temperature are slightly larger than the slopes of calculated values.

**B. Cubic and hexagonal diamond**

By using the same method as we used for hexagonal beryllium, the lattice geometry and TDEC for cubic diamond and hexagonal diamond have been calculated, and the results of elastic constants are shown in Figure 3a and 3b, respectively. For hexagonal diamond, five deformation tensors used are the same as for hexagonal beryllium as listed in Table 1, but for cubic diamond, three deformation tensors are used, and they are listed in Table 1. As shown in Figure 3b and Table 2, the calculated TDEC agree well with the experimental values from Migliori et al [40] and Zouboulis et al's experimental works [41]. This further demonstrates the efficiency and excellent accuracy of our computational method. Similar to the results for hexagonal beryllium, the calculated values are slightly larger than experimental ones due to the existence of defects in the real diamond.

In addition, the elastic constants for cubic diamond at 300 K and 1500 K have been calculated by using the FPMD method, and they are plotted in the Figure 3b and also listed in Table 2 by comparing with the results from the QHA method and experimental results. As can been seen from Figure 3b and Table 2, the calculated TDEC at 300 K and 1500 K are in reasonable agreement with the results obtained by the QHA method and experimental results.

## IV. Summary and conclusions

In conclusion, we have proposed a new method for calculating TDEC for crystals with arbitrary symmetry by combining continuum elasticity theory and first principles. By deriving a Gibbs energy function with one variable with respect to strain at given temperature and pressure, a full minimization of the free energy with respect to temperature and lattice parameters can be put into effective operation by performing first principles calculations. Therefore, anisotropic thermal expansion and equilibrium crystal lattice parameters at a given temperature and pressure condition can be calculated. By polynomial fitting the Helmholtz free energy to strain, the second order coefficient can be obtained. It equals to the corresponding linear combination of elastic constants. By solving simultaneously the corresponding linear equations for the second order coefficient and the corresponding linear combination of elastic constants, the TDEC can be calculated. We have also applied our method to cubic diamond, hexagonal diamond and beryllium crystals. Our calculated results agree well with experimental ones, demonstrating high accuracy of the presented method.

## Acknowledgments

This work was supported by the National Natural Science Foundation of China (Grant No.’s 51121061, 51131002) and the Program for New Century Excellent Talents in Universities of China (NCET-07-0139). R. M. acknowledges the support from the NSERC and CRC programs, Canada. The authors also acknowledge the staff of the Center for Computational Materials Science, Institute for Materials Research, Tohoku University, for computer use.

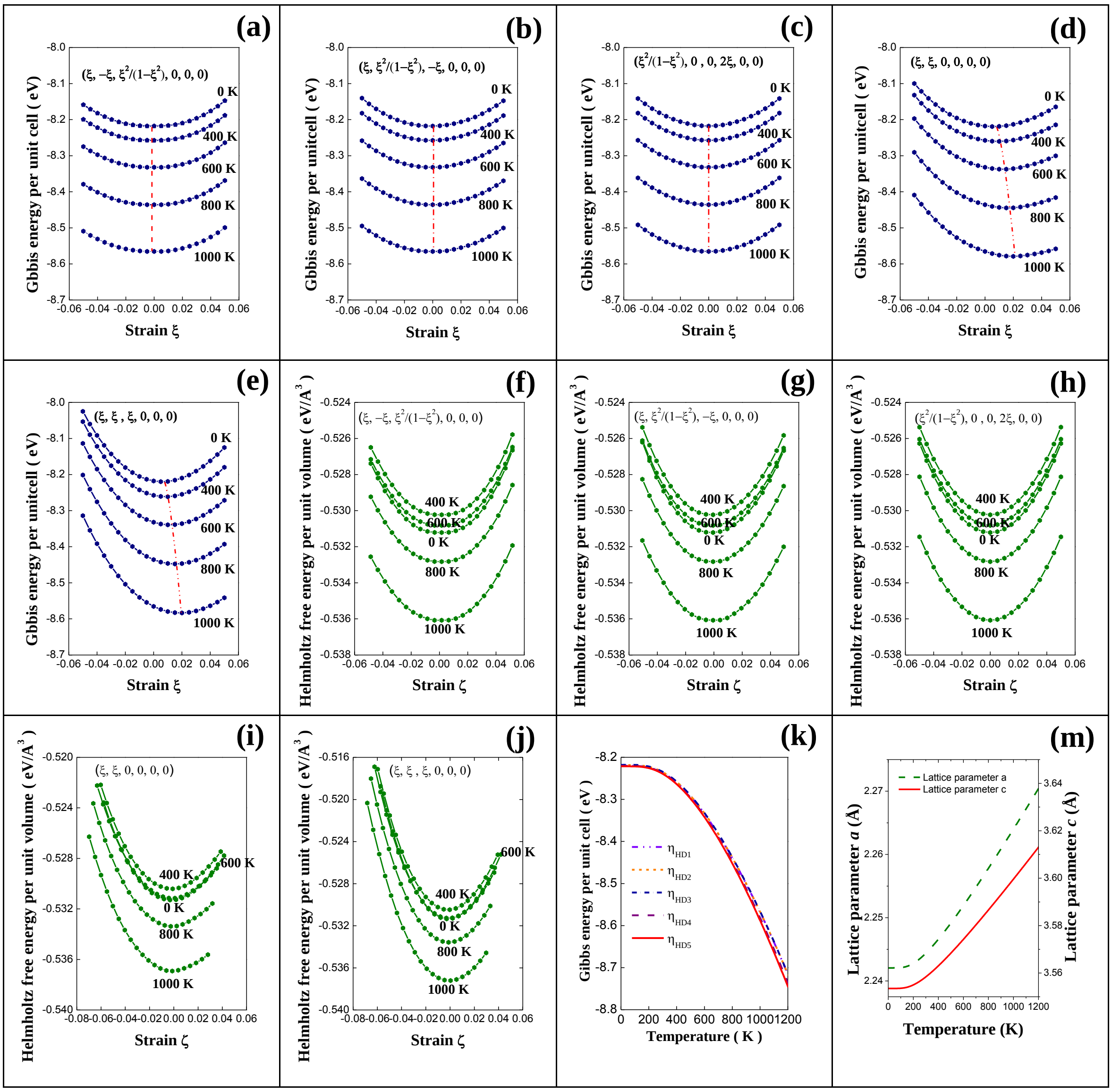

**Figure 1** Calculated results for hexagonal beryllium. (a)~(e) Gibbs energy as a function of strain for different deformation tensors. (f)~(j) Recalculated relationships between Helmholtz free energy density and strain with respect to different deformation tensors. (k) Equilibrium lattice Gibbs energy against temperature with respect to five deformation tensors. (m) Calculated temperature dependent lattice parameter a and c for hexagonal beryllium.

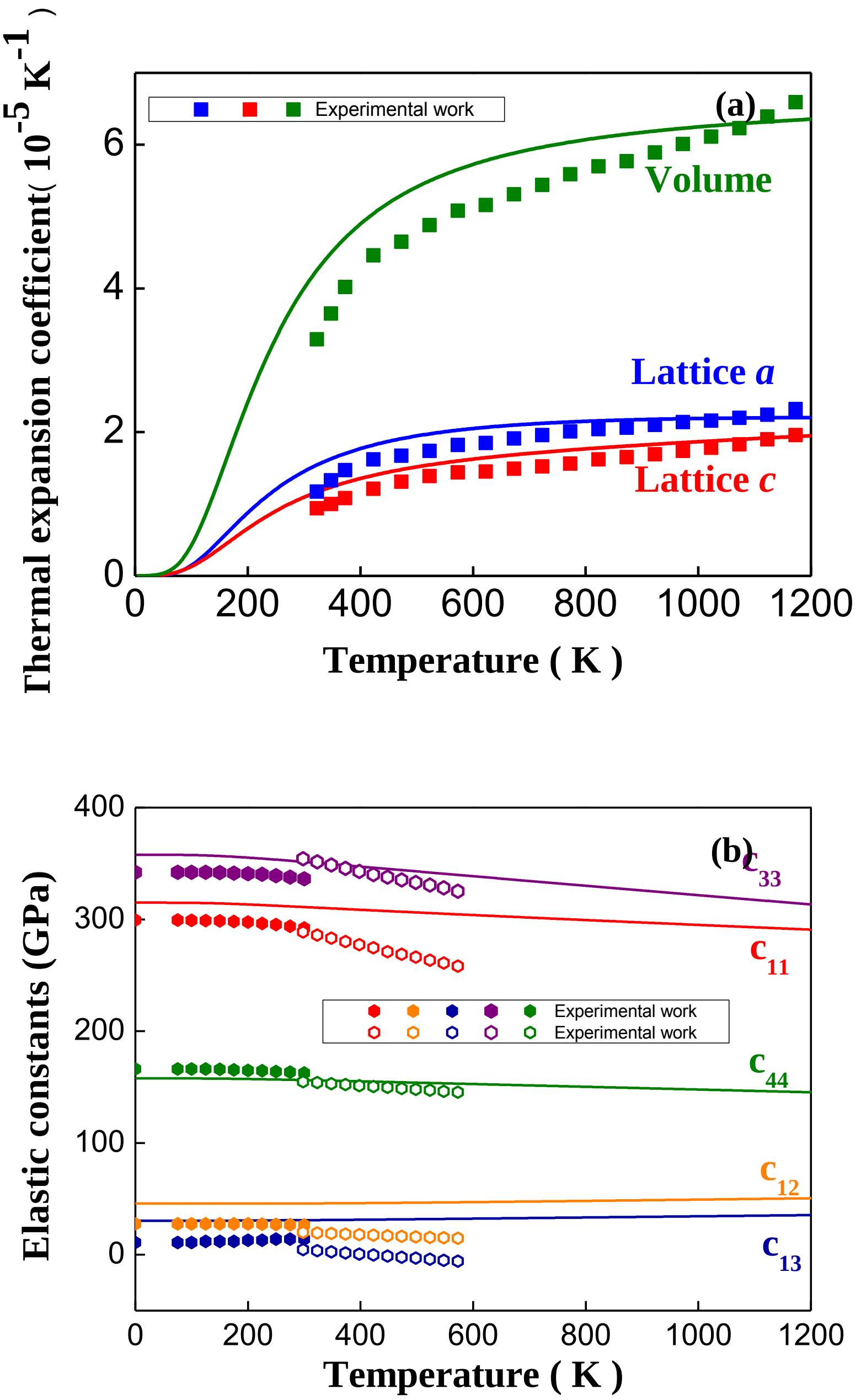

**Figure 2** (a) Comparison of calculated and experimental anisotropic thermal expansion coefficients for hexagonal beryllium. Solid lines are our calculated results, and solid cubics are experimental results from Ref. [37]. (b) Comparison of calculated and experimental temperature dependent elastic constants for hexagonal beryllium. Solid lines are our calculated results, solid hexagons are experimental results from Ref. [38], open hexagons are experimental results from Ref. [39].

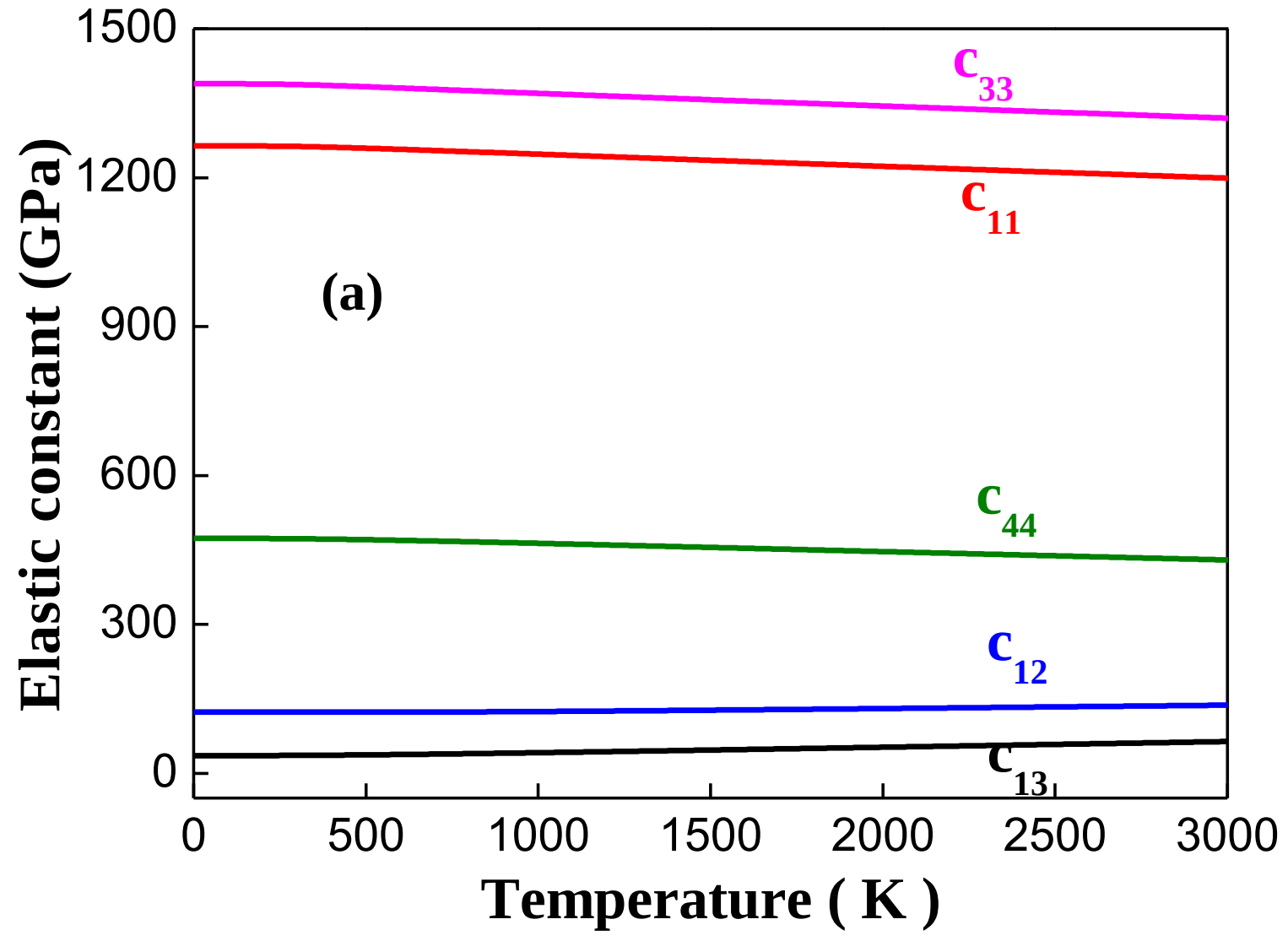

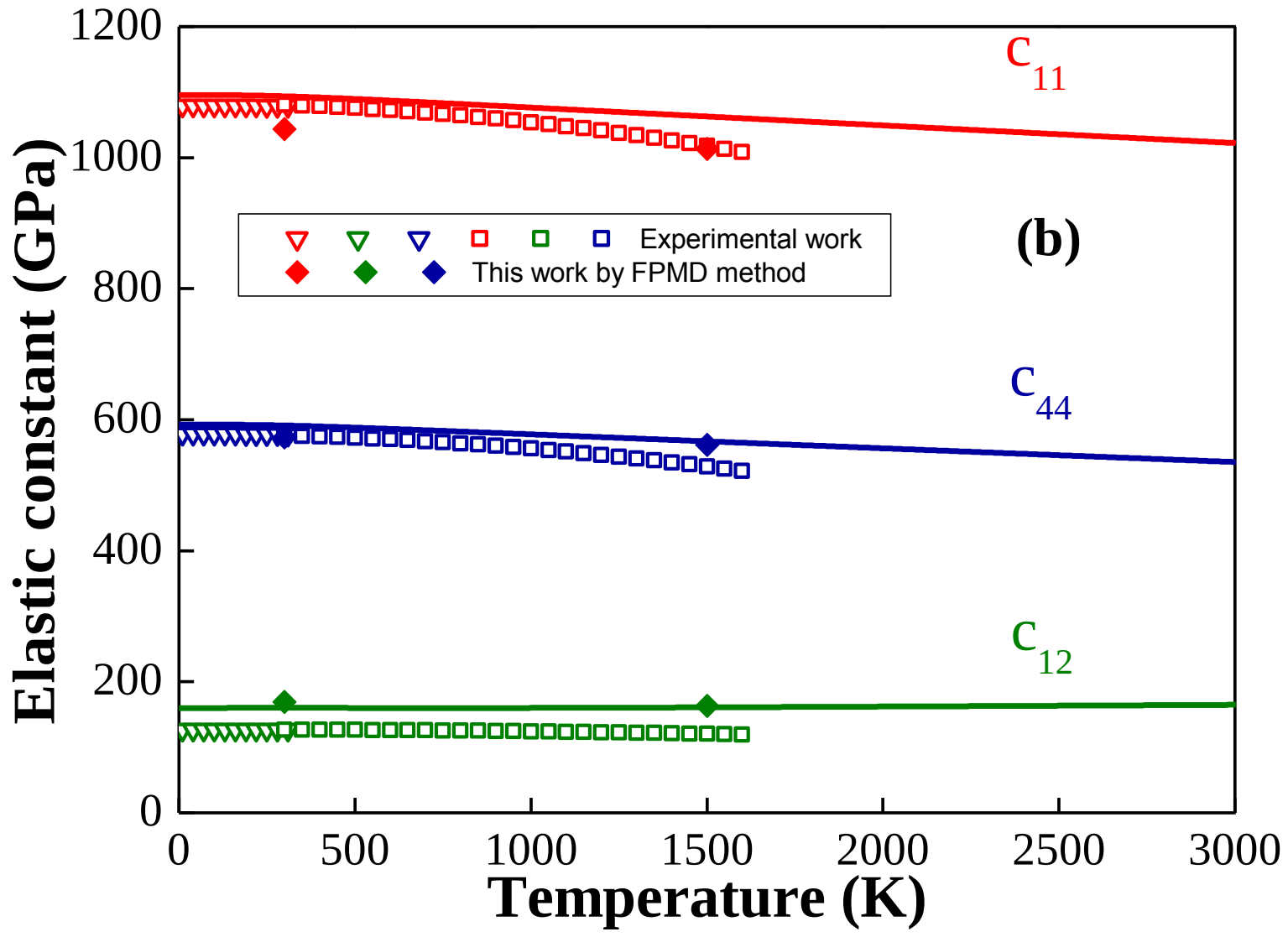

**Figure 3** (a) Calculated temperature dependent elastic constants for hexagonal diamond. (b) Comparison of calculated and experimental temperature dependent elastic constants for cubic diamond. Lines are present calculated results by using the QHA method, inverted triangles are experimental results from [40], cubics are experimental results from [41].

**Table 1**. Deformation tensors used to calculate the elastic constants for cubic lattice and hexagonal lattice. Linear combinations of elastic constants (LCEC) equal to second-order strain derivatives of the Helmholtz free energy under the corresponding deformation modes (or deformation tensors).

| Type of crystals lattice | Deformation tensors | LCEC |
|---|---|---|
| Cubic lattice | $\xi_1 = (\xi - 1, -\xi + 1, \frac{(\xi-1)^2}{1-(\xi-1)^2}, 0, 0, 0)$ | $2C_{11}$-$2C_{12}$ |
| | $\xi_2 = (\frac{(\xi-1)^2}{1-(\xi-1)^2}, 0, 0, 2(\xi-1), 0, 0)$ | $4C_{44}$ |
| | $\xi_3 = (\xi - 1, \xi - 1, \xi - 1, 0, 0, 0)$ | $3C_{11}$+$6C_{12}$ |
| Hexagonal lattice | $\xi_1 = (\xi - 1, -\xi + 1, \frac{(\xi-1)^2}{1-(\xi-1)^2}, 0, 0, 0)$ | $2C_{11}$-$2C_{12}$ |
| | $\xi_2 = (\xi - 1, \frac{(\xi-1)^2}{1-(\xi-1)^2}, -\xi + 1, 0, 0, 0)$ | $C_{11}$+$C_{33}$-2 $C_{13}$ |
| | $\xi_3 = (\frac{(\xi-1)^2}{1-(\xi-1)^2}, 0, 0, 2(\xi-1), 0, 0)$ | 4 $C_{33}$ |
| | $\xi_4 = (\xi - 1, \xi - 1, 0, 0, 0, 0)$ | $2C_{11}$+2 $C_{12}$ |
| | $\xi_5 = (\xi - 1, \xi - 1, \xi - 1, 0, 0, 0)$ | $2C_{11}$+$2C_{12}$+$4C_{13}$+$C_{33}$ |

**Table 2**. Comparation of elastic constant of cubic diamond at 300 and 1500 K from the QHA and FPMD with experimental results.

| Elastic Constants | T = 300 K | | | | T = 1500 K | |
|---|---|---|---|---|---|---|
| | Experimental data by Migliori et al [40] | Experimental data by Zouboulis et al [41] | This work by QHA | This work by FPMD | This work by QHA | This work by FPMD |
| $C_{11}$ | 1078.3 | 1080.4 | 1093.7 | 1043.7 | 1062.9 | 1013.5 |
| $C_{12}$ | 126.5 | 127.0 | 160.2 | 169.1 | 160.9 | 163.3 |
| $C_{44}$ | 577.5 | 576.0 | 591.2 | 577.5 | 567.0 | 561.7 |